\documentclass[12pt]{article}

\usepackage{amssymb}
\usepackage{amsmath}
\usepackage{latexsym}
\usepackage{yfonts}

\catcode`@=11
\renewcommand{\section}{\@startsection{section}{1}{0pt}{\medskipamount}
{\medskipamount}{\large\bf}}
\numberwithin{equation}{section}
\catcode`@=12

\def\beq{\begin{eqnarray}}    
\def\eeq{\end{eqnarray}}      

\def\ln{\,\mbox{ln}\,}                  
\def\tr{\,\mbox{tr}\,}                  
\def\sTr{\,\mbox{sTr}\,}                

\def\pa{\partial}                       

\def\={\ =\ }

\begin{document}

\begin{center}

{\Large\bf A systematic study of finite field dependent BRST-BV
transformations in $Sp(2)$ extended field-antifield formalism}

\vspace{18mm}

{\Large Igor A. Batalin$^{(a)}\footnote{E-mail: batalin@lpi.ru}$\;,
Klaus Bering$^{(b)}\footnote{E-mail:bering@physics.muni.cz}$\;,
Peter M. Lavrov$^{(c, d)}\footnote{E-mail:
lavrov@tspu.edu.ru}$\;, \\ Igor V. Tyutin$^{(a)}\footnote{E-mail:
tyutin@lpi.ru}$ }

\vspace{8mm}

\noindent ${{}^{(a)}}$
{\em P.N. Lebedev Physical Institute,\\
Leninsky Prospect \ 53, 119 991 Moscow, Russia}

\noindent ${{}^{(b)}}$
{\em Masaryk University,  Faculty  of  Science,\\
Kotlarska  2,  611 37  Brno, Czech Republic}

\noindent  ${{}^{(c)}}
${\em
Tomsk State Pedagogical University,\\
Kievskaya St.\ 60, 634061 Tomsk, Russia}

\noindent  ${{}^{(d)}}
${\em
National Research Tomsk State  University,\\
Lenin Av.\ 36, 634050 Tomsk, Russia}

\vspace{20mm}

\begin{abstract}
\noindent In the framework of $Sp(2)$ extended Lagrangian
field-antifield BV formalism we study systematically the role of
finite field-dependent BRST-BV transformations. We have proved that
the Jacobian of a finite BRST-BV transformation is capable of
generating arbitrary finite change of the gauge-fixing function in
the path integral.
\end{abstract}

25.07.2014

\end{center}

\vfill

\\

\section{INTRODUCTION AND SUMMARY}

In the previous articles \cite{BLT-FFDT-BFV,
BLT-FFDT-BV,BLT-FFDT-BFV-Sp2} systematic studies of the role of
finite field dependent BRST transformations have been conducted in
various types of quantization formalisms. Firstly, this was done in
the framework of the standard generalized Hamiltonian BFV formalism
\cite{BLT-FFDT-BFV}, as well as
 in its $Sp(2)$ extended  counterpart \cite{BLT-FFDT-BFV-Sp2}.
 Secondly, a similar
study was performed   in the framework of the standard Lagrangian
field-antifield BV formalism \cite{BLT-FFDT-BV}.  The main result in
all cases is that the Jacobian of a finite field dependent BRST
transformation is capable of reproducing an arbitrary finite change
of the gauge-fixing function in the corresponding path integral.

In the present article, we will develop a similar study in the
$Sp(2)$ extended Lagrangian field-antifield formalism
\cite{BLTl1,BLTl2,BLTl3} (see also an alternative approach to the
problem of BRST-antiBRST invariant quantization of general gauge
theories in Lagrangian formalism \cite{Hull}).  The main new
feature, in the situation at hand, is that the algebra of BRST-BV
transformations is open in the antifield sector, while it is Abelian
in the sector of fields.  It is therefore  in general  impossible to
 have integrable partial
differential  equations in the $Sp(2)$ extended BRST parameters for
the total set of field-antifield variables. However, as the
parameters are only allowed to depend on fields, the corresponding
partial  differential  equations are integrable. At the same time,
for the total set of variables, one can formulate the Lie equations
in terms of a Bosonic rescaling variable for the Fermionic BRST
parameters. Then, when considering the Jacobian of the
transformation, we  can  split the Jacobian into two pieces: the
first one is calculated for constant BRST parameters; the second
piece comes from the field dependence of the BRST parameters. The
second piece is just responsible for generating the required change
of  the gauge-fixing function  in the path integral. It appears that
integrable Lie equations in the field sector are consistent with the
compensation  equation, which determines  the necessary field
dependence for the BRST parameters.
\\

\section{STANDARD  FORMULATION  AND  $Sp(2)$  DIFFERENTIAL  FOR  FIELDS}

Let
\beq
\label{E2.1}
{ \Phi^{A} ,\; \pi^{Aa}, \;\Phi^*_{Aa}, \; \Phi^{**}_{A} }         
\eeq be a set of variables necessary to describe an arbitrary
Lagrangian gauge field theory within the framework of $Sp(2)$
extended field-antifield formalism \cite{BLTl1,BLTl2,BLTl3}. Here in
(\ref{E2.1}) a capital Latin index  $"A"$, $"B"$, $\ldots$, from the
beginning of the alphabet,  should be understood in the sense of
DeWitt's condensed notation  \cite{DeWitt} as a condensed index of
fields and antifields, while a small Latin index $"a"$, $"b"$,
$\ldots$, from the beginning of the alphabet,  is a vector index
of the $Sp(2)$ group. The variables
(\ref{E2.1}) have Grassmann parities \beq \label{E2.2} {
\varepsilon_{A}, \;\;\varepsilon_{A} + 1,\;\; \varepsilon_{A} +
1,\;\;
 \varepsilon_{A} },
 \eeq                       
respectively.
We will denote partial derivatives with respect to the variables (\ref{E2.1}) as
\beq
\label{E2.3}
{  \pa_{A},\; \pa_{Aa}, \;\pa^{Aa},\; \pa^{A} }.      
\eeq
In terms of the variables (\ref{E2.1}), the path integral for the
partition function reads
\beq
\label{E2.4}
Z_{F}=\int D\Phi D\pi D\Phi^* D\Phi^{**} D\lambda \exp\{ ( i / \hbar ) W_{F} \},
\eeq
where
\beq
\label{E2.5}
W_{F}=W + \Phi^*_{Aa} \pi^{Aa}
+ ( \Phi^{**}_{A} - F \overleftarrow{\pa}_{A} )\lambda^{A} +
 (1/2)F \overleftarrow{\pa}_{A} \pi^{Aa} \overleftarrow{\pa}_{B}\pi^{Bb}
\varepsilon_{ba},                       
\eeq
 and where $\varepsilon_{ab}=-\varepsilon_{ba}$ is the constant $Sp(2)$
invariant tensor, while $\varepsilon^{ab}=-\varepsilon^{ba}$ stands for its
inverse.  In (\ref{E2.4}) we have also integrated
over dynamically passive Lagrange
multipliers $\lambda^{A}$ with Grassmann parity  $\varepsilon_{A}$.
 The gauge-fixing  function $F( \Phi )$ in (\ref{E2.5})
is a Boson.
 The quantum  master action
$W=W( \Phi, \Phi^*, \Phi^{**} )$  satisfies the $Sp(2)$ extended quantum master
equation,
\beq
\label{E2.6}
( \Delta^{a} + ( i / \hbar ) V^{a} ) \exp\{ ( i / \hbar) W \} = 0,    
\eeq
where
\beq
\label{E2.7}
&&\Delta^a = (-1)^{\varepsilon_A}\pa_A\,\pa^{Aa},  \\          
\label{E2.8}
&&V^{a} = \varepsilon^{ab}\,\Phi^*_{Ab}\,\pa^{A}.              
\eeq
Fermionic operators (\ref{E2.7}), (\ref{E2.8}) satisfy the
algebra\footnote{Here $[,]$ means the supercommutator, which is
defined for any quantities $G,H$ as $[G,H]=GH-(-1)^{\varepsilon(G)\varepsilon(H)}HG$.}
\beq
\label{E2.9}
&& [ \Delta^{a}, \Delta^{b} ] = \Delta^{ \{a } \Delta^{ b\} } = 0, \\     
\label{E2.10}
&&[ V^{a}, V^{b} ] = V^{ \{a } V^{ b\} } = 0,   \\        
\label{E2.11}
&&[ \Delta^{ \{a }, V^{ b\} } ] = 0.          
\eeq
Here and below $...\{a ... b\}...$ means symmetrization in $a$ and $b$ indices.
 The quantum master equation (\ref{E2.6}) can equivalently be rewritten
in its quadratic form, which is convenient for $\hbar$ series expansion,
\beq
\label{E2.12}
(1/2) ( W, W )^{a} + V ^{a} W = i \hbar \Delta^{a} W,          
\eeq
where in the l. h. s., we have denoted the $Sp(2)$ extended antibracket
\beq
\label{E2.13}
( F, G )^{a}
= F (\overleftarrow{\pa}_{A} \pa^{Aa} -\overleftarrow{\pa}^{Aa} \pa_{A} ) G =
- ( G, F )^{a} (-1)^{ ( \varepsilon(F) + 1)( \varepsilon(G) + 1)}.     
\eeq
The antibracket (\ref{E2.13}) is  Grassmann  odd,
\beq
\label{E2.14}
\varepsilon( ( F, G )^{a} )
= \varepsilon( F ) + \varepsilon( G ) + 1;  
\eeq
it satisfies the Leibniz rule,
\beq
\label{E2.15}
( F, G H )^{a} = ( F, G )^{a} H + G ( F, H )^{a}
(-1)^{ ( \varepsilon( F ) + 1) \varepsilon(G)};          
\eeq
it satisfies the Jacobi identity,
\beq
\label{E2.16}
( ( F, G )^{ \{a }, H )^{ b\} } (-1)^{ ( \varepsilon( F ) + 1 ) ( \varepsilon( H ) + 1 ) }
+{\rm cycle}( F, G, H ) = 0;                   
\eeq
and  it is differentiated by both the operators
(\ref{E2.7}) and (\ref{E2.8}),
\beq
\label{E2.17}
&&\Delta^{ \{a } ( F, G )^{ b\} }  = ( \Delta^{ \{a } F, G )^{ b\} } -
 ( F, \Delta^{ \{a } G )^{ b\} } (-1)^{ \varepsilon( F ) },\\        
\label{E2.18} &&V^{ \{a } ( F , G )^{ b\} } = ( V ^{ \{a } F, G )^{
b\} } -  ( F, V^{ \{a } G )^{ b\} } (-1)^{ \varepsilon( F) }.        
\eeq
Now let us make the essential observation that the
gauge-fixed quantum action can be rewritten in the following equivalent form
\beq
\label{E2.19} W_{F} = W + \Phi^*_{Aa } \pi^{Aa} +
\Phi^{**}_{A} \lambda^{A} +
(1/2) F \overleftarrow{d^{a}d^{b}} \varepsilon_{ba},           
\eeq
where the  (right) Fermionic $Sp(2)$ differential is defined
by
\beq
\label{E2.20}
&&\overleftarrow{d^a} = \overleftarrow{\pa}_{A} \pi^{Aa}-
\overleftarrow{\pa}_{Ab} \lambda^{A} \varepsilon^{ba},    \\      
\label{E2.21}
&&[ \overleftarrow{d^a, d^b} ] =
\overleftarrow{d^{ \{a } d^{ b\}}}=0,    \\     
\label{E2.22}
&&\overleftarrow{d^a d^b d^c} = 0.
\eeq
Its counterpart acting  from the left  reads
\beq
\label{E2.23}
&&d^{a} = (-1)^{\varepsilon_{A} + 1}\pi^{Aa} \pa_{A}
+ (-1)^{\varepsilon_{A}}\lambda^{A} \varepsilon^{ab} \pa_{Ab}, \\       
\label{E2.24}
&&[d^a,d^b] = d^{\{a} d^{ b\} } = 0 , \quad  d^ad^bd^c = 0,\\   
\label{E2.25}
&&G\overleftarrow{d^a}\mu_a=\mu_ad^a G,\quad
\varepsilon(\mu_a)=1,
\eeq
where $G$ is a function of all variables.
It is a remarkable feature of (\ref{E2.19}) that the dependence on
the gauge-fixing function $F$ is  here accumulated  in the fourth
term alone, which is similar to  what happens in the corresponding
Hamiltonian formulation \cite{BLT-FFDT-BFV-Sp2}.
\\

\section{$Sp(2)$ EXTENDED BRST-BV GENERATORS AND \\ THEIR  ALGEBRA}

 Now let us define the (left)
$Sp(2)$ extended BRST-BV generators by
\beq
\label{E3.1}
\textfrak{D}^a = d^{a} + ( W \overleftarrow{\pa}_{A} ) \pa^{Aa} +
\varepsilon^{ab} \Phi^*_{Ab} \pa^{A}.        
\eeq
Due to the quantum master equation (\ref{E2.12}), it follows from (\ref{E3.1})
that
\beq
\label{E3.2}
\textfrak{D}^a W_{F} = i \hbar \Delta^{a} W_{F},         
\eeq
and  that  the generators (\ref{E3.1}) satisfy the algebra
\beq
\label{E3.3}
[ \textfrak{D}^a, \textfrak{D}^b] =
(D^{\{a} W \overleftarrow{\pa}_{A}) \pa^{ Ab\} }+
W\overleftarrow{\pa^{ A \{a }} (\pa_{A} W \overleftarrow{\pa}_{B}) \pa^{B b\}} ,
\eeq
where
\beq
\label{E3.4}
&&D^{a} = d^{a} + i \hbar \Delta^{a},      \\       
\label{E3.5}
&&[ D^{a}, D^{b} ] = D^{ \{a } D^{ b\} } = 0, \;\; [D^a,\pa_A]=0.   
\eeq
The (right) $Sp(2)$ extended BRST-BV generators are written as
\beq
\overleftarrow{\textfrak{D}^a}=\overleftarrow{d^a}+\overleftarrow{\pa}^{Aa}(\pa_A W)+
\overleftarrow{\pa}^A\Phi^*_{Ab}\varepsilon^{ba}(-1)^{\varepsilon_A},
\eeq
so that the relation
\beq
G\overleftarrow{\textfrak{D}^a}\mu_a=\mu_a\textfrak{D}^aG, \qquad
\varepsilon(\mu_a)=1
\eeq
holds for a function $G$ of all variables.

Thus, we
conclude from (\ref{E3.3}) that in the total field-antifield space,
the algebra is open, while  it is Abelian in the field sector $\{
\Phi^{A}, \pi^{Aa} \}$.
\\

\section{LIE EQUATIONS AND FINITE $Sp(2)$ EXTENDED BRST-BV TRANSFORMATIONS}

Let us denote the set of field-antifield variables (\ref{E2.1}) by
$z^{i}$. We next consider the Lie equation for a finite BRST-BV transformation,
restricted to a one-parameter subgroup of $t$-rescaling of the Fermionic
parameter $\mu_{a}, \mu_{a} \rightarrow t \mu_{a}$,
where $t$ is a Bosonic parameter,
\beq
\label{E4.1}
&&\frac{d }{ dt }
\overline{z}^i(z,t\mu) = \mu_{a} \overline{\textfrak{D}^a z^{i}} =
 \overline{z^{i} \overleftarrow{\textfrak{D}^{a}}}  \mu_{a},   \\    
\label{E4.2}
&&\overline{z}^{i} ( t = 0 ) = z^{i}.            
\eeq
For constant $\mu_{a}$, the formal solution to that equations reads
\beq
\label{E4.3}
\overline{z}^{i}  =  \exp\{t \mu_{a} \textfrak{D}^{a} \} z^{i}  =
z^{i} \exp\{t \overleftarrow{\textfrak{D}^{a}} \mu_{a} \}.        
\eeq

It is also natural to extend the solution (\ref{E4.3}) to the case of
parameters $\mu_a$ being dependent of the initial data (\ref{E4.2}),
$\mu_a=\mu_a(z)$, as follows: all $\mu_a$ should stand to the left (right) of
all $\textfrak{D}^a$ in the first (second) equality in (\ref{E4.3}).  We will
mean that extension when considering the Jacobian and the compensation
equation in Sec. 5.

As the algebra (\ref{E3.3}) is open in the total field-antifield space, it is
impossible to deduce integrable partial  differential
equations in the parameters $\mu_{a}$ themselves for all the variables
$z^{i}$. However, if one considers only the field sector
$z^{\alpha} = \{ \Phi^{A},\pi^{Aa} \}$, then the integrable partial
 differential equations
\beq
\label{E4.4}
\pa^{a} \overline{z}^{\alpha}(z,t\mu) = t \overline{d^{a}z^{\alpha}},\quad
 \pa^{a} = \frac{\pa}{\pa \mu_a},              
\eeq
directly imply the corresponding equations (\ref{E4.1}) in the field sector
$z^{\alpha}$. These equations (\ref{E4.4} ) are integrable due to (\ref{E2.24}).
When multiplied by $\mu_{a}$ from the left,
equations (\ref{E4.4}) in the $z^{\alpha}$ sector
yield exactly (\ref{E4.1}), cf.\ eq.\ (\ref{E3.1}).
A version of (\ref{E4.4}) using right differentials  reads
\beq
\label{E4.5}
\overline{z}^{\alpha}(z,t\mu)\overleftarrow{\pa}^{a} =
\overline{z^{\alpha}\overleftarrow{d^{a}}} t.                
\eeq
For further convenience, let us also write down a counterpart to (\ref{E4.5})
for inverse transformation,
\beq
\label{E4.6}
z^{\alpha}(\overline{z},t\mu) \overleftarrow{\pa}^{a} =
- z^{\alpha} \overleftarrow{d^{a}} t.                              
\eeq
It follows from (\ref{E3.2}) and (\ref{E4.1}) that
\beq
\label{E4.7}
(d / dt) \overline{W_{F} } =
\mu_{a} i \hbar\overline{\Delta}^{a}\;\overline{W_{F}}.   
\eeq

\section{JACOBIAN  AND  COMPENSATION  EQUATION }

Let again $z^{i}$  be the set (\ref{E2.1}) of
field-antifield variables. Now we  would like
to study the general structure of the following Jacobian
\beq
\label{E5.1}
\ln J = \sTr \ln [ ( \overline{z}^{i} \overleftarrow{\pa}_{k} ) ],    
\eeq
where $\overline{z}^{i}$ is defined in (\ref{E4.1}) with parameters
$\mu_{a}$ depending on the field  variables
$z^{\alpha} = \{ \Phi^{A}, \pi^{Aa}\}$ and $\lambda^A$.
In general, we have
\beq
\label{E5.2}
\ln J=\sTr\ln[(\overline{z}^i(z,t\mu)\overleftarrow{\pa}_{k})]+
\sTr\ln[\delta^{\alpha}_{\beta} -
(z^{\alpha}(\overline{z},t\mu)\overleftarrow{\pa}^{a})
( \mu_{a} \overleftarrow{\pa}_{\beta})].                 
\eeq
Here  on  the r. h. s., in the first term, the
 right
$\overleftarrow{\pa}_{k}$ derivative is taken  for  constant $\mu$,
while the second term is just an effect of $\mu$ being $z^{\alpha}$ dependent.
Then, by using (\ref{E4.1}), (\ref{E4.6}) we get
\beq
\label{E5.3}
( d / dt ) \ln J =
\mu_{a} \overline{\Delta^{a}} \;\overline{ W_{F} } -
\tr [ ( 1 + t \kappa )^{-1} \kappa ],         
\eeq
where
\beq
\label{E5.4}
\kappa_{a}^{b} = ( \mu_{a} \overleftarrow{d}^{b} ),    
\eeq
and where  we have used that the $t$-derivative of the
second term in (\ref{E5.2}) equals
\beq \label{E5.5} G^{\alpha}_{
\beta} (z^{\beta} \overleftarrow{d}^{a} )
( \mu_{a} \overleftarrow{\pa}_{\alpha} ) (-1)^{\varepsilon_{\alpha}},       
\eeq
with $G^{\alpha}_{\beta}$ satisfying the equation
\beq
\label{E5.6}
G^{\alpha}_{\gamma} + t ( z^{\alpha}\overleftarrow{d}^{a} )
( \mu_{a}\overleftarrow{\pa}_{\beta} ) G^{\beta}_{\gamma}
= \delta^{\alpha}_{\gamma}. 
\eeq
It follows from (\ref{E5.6}) that
\beq
\label{E5.7}
( \mu_{a}\overleftarrow{\pa}_{\alpha} ) G^{\alpha}_{\beta}  =
[ ( 1 + t \kappa )^{-1} ]_{a}^{b} \;( \mu_{b}\overleftarrow{\pa}_{\beta} ). 
\eeq
By substituting  (\ref{E5.7})  into (\ref{E5.5}), we get
for the latter exactly the second term in (\ref{E5.3}).
Now, it follows from (\ref{E4.7}) and (\ref{E5.3}) that
\beq
\label{E5.8}
( d / dt ) ( ( i / \hbar)\overline{W_{F} } + \ln J ) =
- \tr [ ( 1 + t \kappa )^{-1} \kappa ],    
\eeq
so that
\beq
\label{E5.9}
( ( i / \hbar )\overline{W_{F} } + \ln J )_{t = 1} =
( i / \hbar ) W_{F} - \tr  \ln ( 1 + \kappa ) .  
\eeq
In order to change the gauge-fixing function
\beq
\label{E5.10}
F \rightarrow F_{1} = F + \delta F            
\eeq
 with a finite amount $\delta F$, we must have
\beq
\label{E5.11}
- \tr \ln ( 1 + \kappa ) =
( i / \hbar) (1/2) \delta F \overleftarrow{{d}^{a}d^{b}}
\varepsilon_{ba} = - \tr x,                       
\eeq
where
\beq
\label{E5.12}
x_{a}^{b} = - ( i / \hbar ) (1/2) \varepsilon_{ac}
\delta F\overleftarrow{d^{c}d^{b}}.   
\eeq
 We must also require the following condition to hold
\beq
\label{E5.13}
\ln ( 1 + \kappa ) = x,       
\eeq
so that
\beq
\label{E5.14}
\kappa = \exp( x ) - 1,                  
\eeq
or explicitly
\beq
\label{E5.15}
\mu_{a}\overleftarrow{d}^{b} = [ \exp( x ) - 1 ]_{a}^{b}.      
\eeq
We call (\ref{E5.15}) "a compensation equation".
 This equation determines
the necessary dependence of the parameters $\mu_{a}$ on fields.
There is an obvious explicit solution to that equation, namely
\beq
\label{E5.16}
\mu_{a} = - ( i / \hbar ) [ f( x ) ]_{a}^{b} ( 1/2) \varepsilon_{bc}
\delta F \overleftarrow{d}^{c},                       
\eeq
where
\beq
\label{E5.17}
f( x ) = ( \exp( x ) - 1 )\, x^{-1}.                   
\eeq
If one chooses the parameters $\mu_{a}$ in the form (\ref{E5.16}),
then it follows due to (\ref{E5.9}) and  (\ref{E5.11}),
 that
\beq
\label{E5.18}
Z_{F_{1}}  =  Z_{F},                         
\eeq
so that the partition function is independent of the gauge-fixing $F$.
\\

 \section{$ Sp(2 )$ EXTENDED  MODIFIED  WARD IDENTITIES}

Let us define the generating functional for Green's functions by adding the
action of external sources to the gauge-fixed quantum  action $W_{F}$
(\ref{E2.5}) or (\ref{E2.19}),
\beq
\nonumber
Z_{F}( \zeta, \zeta^*, \zeta^{**} ) &=&
\int  D\Phi D\pi  D\Phi^* D\Phi^{**} D\lambda
\exp\{ ( i  / \hbar ) [ W_{F} + \zeta_{\alpha} z^{\alpha} +\\
\label{E6.1}
&&\qquad\qquad+\zeta^*_{\alpha a}  z^{\alpha} \overleftarrow{d^{a}}  +
\zeta^{**}_{\alpha}  z^{\alpha }\overleftarrow{d^{a}d^{b}} (1/2)
\varepsilon_{ba} ] \},                                
\eeq
where, again,  $z^{\alpha} = \{ \Phi^{A}, \pi^{Aa} \}$ is the field sector.

Let us make in (\ref{E6.1}) an infinitesimal change with constant $\mu_{a}$,
\beq
\label{E6.2}
\delta z^{i} = z^{i}
\overleftarrow{\textfrak{D}^{a}}
\mu_{a},  \quad
\mu_{a}\rightarrow  0,                   
\eeq
of the total set $z^{i}$ of the field-antifield variables (\ref{E2.1}).
We get then the standard Ward Identities
\beq
\label{E6.3}
( \zeta_{\alpha} \pa^{\alpha a} +
\varepsilon^{ab} \zeta^*_{\alpha b} \pa^{\alpha} ) Z_{F} = 0,     
\eeq
where $\pa^{\alpha}$ and $\pa^{\alpha a}$ is the partial derivative with respect
to $\zeta^{**}_{\alpha}$ and $\zeta^*_{\alpha a}$, respectively.

Now let us perform in (\ref{E6.1}) a finite BRST-BV transformation
defined by (\ref{E4.1}) and   (\ref{E4.2}),
with arbitrary $\mu_{a}( z )$;
we get then what we call "a modified Ward identity",
\beq
\nonumber
&&<\exp\{ ( i / \hbar ) [ \zeta_{\alpha} (\overline{z} - z )^{\alpha} +
 \zeta^*_{\alpha a } (\overline{z^{\alpha}\overleftarrow{d^{a}}} - z^{\alpha}
\overleftarrow{d^{a}} ) +
\zeta^{**}_{\alpha} (\overline{z^{\alpha}\overleftarrow{d^{a}d^{b}} } -
z^{\alpha} \overleftarrow{d^{a}d^{b}} ) (1/2) \varepsilon_{ba} ] -\\
\label{E6.4}
&&-\tr \ln ( \delta_{a}^{b}  + \mu_{a}\overleftarrow{d^{b}}) \} >_{F; \zeta,
\zeta^*, \zeta^{**}} = 1,                
\eeq
where we have defined the source-dependent quantum mean value
\beq
\nonumber
&&< ( \ldots ) >_{F; \zeta, \zeta^*, \zeta^{**}} =
[ Z_{F}( \zeta, \zeta^*, \zeta^{**} )]^{-1}
\int D\Phi D\pi  D\Phi^* D\Phi^{**} D\lambda  ( \ldots )
\exp\{ ( i / \hbar )[ W_{F} + \\
\label{E6.5}
&&+\zeta_{\alpha} z^{\alpha}+\zeta^*_{\alpha a}  z^{\alpha}\overleftarrow{d^{a}}  +
\zeta^{**}_{\alpha}  z^{\alpha}\overleftarrow{d^{a}d^{b}} (1/2)
\varepsilon_{ba} ] \} .                    
\eeq Now, let us choose in (\ref{E6.4}) the parameters $\mu_{a}$ to
coincide with the solution (\ref{E5.16}) to the compensation
equation with the inverse sign of $\delta F$; then we get from
(\ref{E6.4}) the relation generalizing  the one (\ref{E5.18}) to the
presence of external sources.
\beq
\label{E6.6}
Z_{F_{1}}  = Z_{F} <
\exp\{ ( i  / \hbar ) [ \ldots ] \} > _{F; \zeta, \zeta^*,
\zeta^{**}},         
\eeq
where $[ \ldots ]$ means the expression in the square brackets in the
exponential  on  the l. h. s.  of  (\ref{E6.4}).
\\

\section*{Acknowledgments}
\noindent
K.B. would like to thank K.P.~Zybin and the Lebedev Physics Institute
for warm hospitality.
The work of I.A.B. is supported in part by the RFBR grants 14-01-00489
and 14-02-01171.
The work of K.B.  is supported by the Grant Agency of the Czech Republic
(GACR) under the grant P201/12/G028.
The work of P.M.L. is partially supported by the Ministry of Education and
Science of Russian Federation, grant TSPU-122, by the Presidential grant
88.2014.2 for LRSS and  by the RFBR grant 13-02-90430-Ukr.
The work of I.V.T. is partially supported by the RFBR grant 14-02-01171.
\\

\end{document}